\documentstyle[11pt,moriond,epsfig]{article}

\def\Journal#1#2#3#4{{#1} {\bf #2}, #3 (#4)}

\def\ApJ{\em Astrophys. J.}
\def\AA{\em Astron. Astrophys.}
\def\ARAA{\em Annual Rev. Astron. Astrophys.}

\def\M{{\cal M}}
\def\R{{\cal R}}
\def\F{{\cal F}}
\def\S{{\cal S}}
\def\D{{\cal D}}

\def\be{\begin{equation}}
\def\ee{\end{equation}}
\def\bea{\begin{eqnarray}}
\def\eea{\end{eqnarray}}

\begin{document}
\vspace*{4cm}
\title{BINARY BLACK HOLE FORMATION AND MERGINGS}

\author{K.A. POSTNOV, M.E. PROKHOROV}

\address{Sternberg Astronomical Institute, 119899 Moscow\\
Russia}

\maketitle\abstracts{The formation and evolution of 
binary black holes (BH) is studied using the modern evolutionary scenario for
very massive stars. Main sequence stars with masses $M>35 M_\odot$ 
are assumed to form a BH in the end of their nuclear evolution. 
The mass of BH formed is parametrized as $M_{bh}=k_{bh}\times M_{SN}$, 
where $M_{SN}$ is the mass of the pre-supernova star taken from evolutionary 
calculations, $k_{bh}\le 1$. The possibility is explored that 
a newly formed BH acquires a kick velocity 0-250 km/s. Binary BH are 
found to merge within the Hubble time at an appreciable rate only
for non-zero kick velocities. We calculate the 
galactic merging rates of binary BH systems, their detection rate 
by the initial laser interferometers, and the distributions
of merging binary BH over orbital eccentricities at different frequencies.
The distribution of angles between BH spins and the orbital angular momentum
is also presented.}

\section{Introduction}

Among the most promising sources of
gravitational waves (GW) which can be detected by the initial LIGO/VIRGO/GEO-600
laser interferometers with a sensitivity of $h_{c}\sim 10^{-21}$ at 
the frequency 100 Hz, binary black hole (BH) systems play a special role.
Their importance stems from their larger masses, in comparison
to binary neutron star (NS) systems. From observations of binary radiopulsars
and studies of X-ray binaries the average mass of a NS is derived to be 
$\approx 1.4 M_\odot$ 
(see \cite{ThorsettChakrabarty98,Taylor99}), 
while the mean mass of BH candidates in X-ray binaries is $\sim 8 M_\odot$ 
(see \cite{Cherepashchuk96,TanakaShibazaki97}). 

Consider a coalescing binary with component masses $M_1$ and $M_2$
lying at a distance $r$. Let the average formation rate per volume
of such binaries be $\F$ and the merging rate within a Hubble time be $\R$, 
$\F\ge \R$. 
As the characteristic GW
amplitude from such a binary at frequency $f$, 
which determines the signal-to-noise ratio ($\S$) at 
the detector when applying a match filtering technique in data analysis, 
is \cite{Thorne87} $h_c\sim \M^{5/6} \nu^{-1/6}r^{-1}$ (here
$\M=(M_1+M_2)^{-1/5}(M_1M_2)^{3/5}$ is the "chirp" mass of the binary), 
the detection rate of such mergings at the detector with a given
$\S$ will be $\D\sim \R \M^{15/6} \S^{-3}$. Thus a strong mass enhancement 
of binary BH detection rate with respect to binary NS detection rate, 
$(\M_{bh}/\M_{ns})^{15/6}$, 
cam
make binary BH mergings even more preferable than binary NS ones
\cite{LPP97a}. 
What is needed is to calculate the formation $\F$ and merging rates $\R$ 
of the corresponding compact binaries. 
This problem has been addressed earlier (e.g. 
\cite{Clark_ea79,TutukovYungelson93},..., see \cite{LPP97b} 
and references therein), 
but due to 
the lack of the accurate knowledge about BH formation parameters
the obtained results have been considered as an order of magnitude
estimates.

Binary NS formation and mergings can be studied both using observations of
binary radiopulsars with secondary NS companion 
\cite{Phinney91,vdHLorimer96,Kalogera99,Prince99} or theoretically
\cite{Narayan_ea91,TutukovYungelson93,LPP97b} etc. Unlike
NS, no binary BH system is (and will be) known until GW observatories are
at work. 

Difficulties in studying BH formation relate not only to the unknown,
from the first principles, mass of the star which leaves a BH in the end of
its evolution and the mass of the formed BH itself, 
but appears already in the description of
the very early evolution of high-mass normal stars because of a significant 
stellar wind mass loss. Until recently, no good evolutionary calculations 
capable of explaining many observational astronomical facts related to massive
star evolution have been available. The situation, however, seems to have
changed after a detailed work of Vanbeveren et al. \cite{Vanbeveren_ea98}.
These new calculations of massive star evolution take into account the 
observed stellar wind mass loss from massive stars, and reproduce Wolf-Rayet
star population characteristics close to those 
observed in the Galaxy and the Magellanic Clouds. 
     
Here we present the results of population synthesis of massive binary
evolution (see \cite{LPP97b} for more detail and references) using these 
new calculations of Vanbeveren et al. \cite{Vanbeveren_ea98}.

\section{Galactic binary BH formation, merging, and detection rates}

\subsection{A new scenario for massive binary evolution}

A spherically-symmetric
mass loss during the main sequence star evolution makes 
the binary semimajor axis $a$ to increase such that after a mass fraction
 $\Delta M$
has been lost from the system $a/a_0=(M_1+M_2)/(M_1+M_2-\Delta M)$
(e.g. van den Heuvel\cite{vdH94}). The process which can  decrease 
the binary 
separation is the tidal interaction between the components 
leading to mass transfer from one star to another. The most violent
mass transfer can result in the common envelope stage during which the 
binary separation can decrease by many times leading sometimes even to the 
coalescence of the components.    
However, an important fact that follows from 
the new evolutionary calculations
\cite{Vanbeveren_ea98} is that primary stars in binaries 
with initial masses $M_1>40 M_\odot$
{\it can not fill} their Roche lobes at all 
because of the strong stellar wind, so it 
essentially evolves like a single star. Of course, in a
small 
fraction of initially very close binaries the more massive primary component
can fill its Roche lobe and mass transfer can begin. But we assumed that 
in {\it all} initial types of binaries the primary with $M_1>40 M_\odot$
does not fill its Roche lobe. This assumption has an advantage of avoiding 
large uncertainties in the evolution with mass transfer (e.g., the degree
of non-conservativeness or the common envelope efficiency). 
As the mass transfer process makes the separation between the
components decrease,
the binary BH merging rates we calculate should be considered as {\it lower
limits}.       

\subsection{BH formation parameters}

At present, there is no full understanding of BH formation. Some
calculations can be found in the literature (see, for example, the recent
paper by Fryer \cite{Fryer99} and references therein), but they 
are not self-consistent. Moreover, the high mass-loss
pertinent to massive star is ignored in these the calculations.
So at this stage it seems justified to approximate BH formation by 
two parameters, the initial mass of a main-sequence star leaving a 
BH in the end of its evolution, $M_{cr}$, and the fraction of the
pre-supernova mass that comes to the BH, $k_{bh}=M_{bh}/M_{SN}$. 
In our calculations we have taken $M_{SN}$ from evolutionary calculations
\cite{Vanbeveren_ea98} and assumed $k_{bh}=0.75$ so that typical BH 
masses lie within the observed range of BH candidate masses 6-12 $M_\odot$.
We have taken $M_{cr}=35 M_\odot$, in accordance with calculations
of \cite{Vanbeveren_ea98,Fryer99}. We did not take into account a possible 
BH formation during matter fall-back onto a NS in core-collapse supernovae
(cf. \cite{Fryer99})
or during hypercritical accretion of matter onto NS in a common envelope
stage \cite{BetheBrown98}. Clearly, these additional channels 
of BH formation will make the formation rate of binary BH systems higher
by some factor that cannot be relaibly etimated at present. Such 
accretion-induced BH have masses not much higher than the limiting NS
mass ($\sim 3 M_\odot$) and from the point of view of GW detection they 
are closer to NS+BH binaries having smaller chirp masses. 

\subsection{Effect of kick velocity}

One of the most important parameters of binary evolution 
is the kick velocity 
acquired by a compact star (a NS or BH). The existence of natal kicks
during NS formation follows from radiopulsar velocity measurements
\cite{Lorimer98} and evolutionary studies of pulsar formation
(e.g., Ref. \cite{PZvdH99}). 
The physical mechanism for the kick velocity is unknown.
The effect of kick velocity on the binary compact star 
formation and coalescence was studied by Lipunov et al. \cite{LPP97b} 
It is usually assumed that the vector of 
kick velocity is arbitrarily oriented in
space and its modulus has some distribution $f(v)$ which we used both in a
Maxwellian form $f(v)_M \propto v^2\exp[-(v/v_0)^2]$ or in the form 
derived from pulsar velocity observations \cite{LPP97b}
$f(v)_{LL}\sim (v/v_0)^{0.19}(1+(v/v_0)^{6.72})^{-1/2}$.

It is unclear now 
whether the kick is associated with BH formation as well. However,
there are astronomical indications that BH candidates (X-ray novae) 
have higher barycentric velocities than ordinary massive binaries
\cite{Cherepashchuk96,jvPWhite96}, 
which can be explained either assuming a substantial mass loss
(which we do not think to occur)
or by a kick velocity of 
$\sim 100$ km/s during BH formation \cite{jvPWhite96}. A long-term 
periodicity observed in some BH candidates can be also due to 
spins of the component being non-parallel to the orbital angular 
momentum. So in our
calculations we assumed BH kick velocities distributed in the same way as 
for NS but with $v_0^{bh}=v_0^{ns}(1-k_{bh})/(1-M_{OV}/M_{SN})$, 
where $k_{bh}=M_{bh}/M_{SN}$ is the fraction of the pre-supernova 
mass $M_{SN}$ collapsing into BH, $M_{OV}=2.5 M_\odot$ is the upper mass limit of NS.
Clearly, for $k_{bh}=1$ no additional velocity is acquired by BH, and when 
$M_{bh}\to M_{OV}$ the kick velocity $v_{bh}\to v_{ns}$.       

The effect of kick velocity is twofold. Depending on the value and 
the kick orientation 
relative to the velocity of the pre-supernova star
before the explosion, it can either disrupt or bind stronger the system
(see e.g. Yamaoka et al.~\cite{Yamaoka_ea93} for more detail). We found 
that for binary BH with typical masses of $7-10 M_\odot$ which form 
during the evolution of massive binaries the kick velocity is mostly 
affects the merging rate. Without kick velocity double BH 
form but most of them recide in too wide binaries to merge in a time
shorter than the Hubble time (cf. \cite{YungelsonPZ98}), so binary BH merging
rate $\R$ is very small to be of significance for GW observations.

\begin{figure}
\hbox to \textwidth{ 
\hbox to 0.5\textwidth{
\psfig{figure=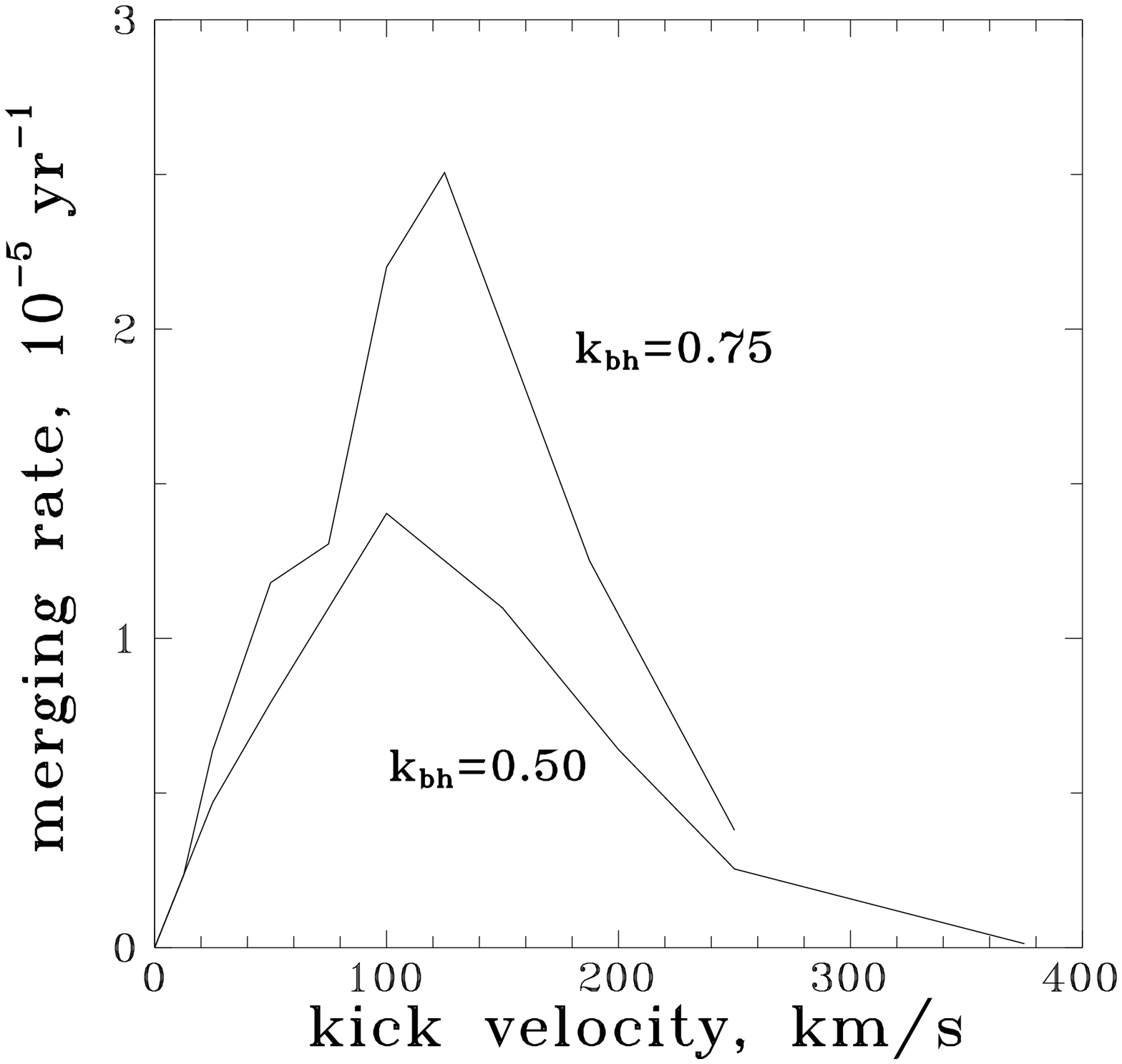,height=2.3in}
}
\hss
\hbox to 0.5\textwidth{
\psfig{figure=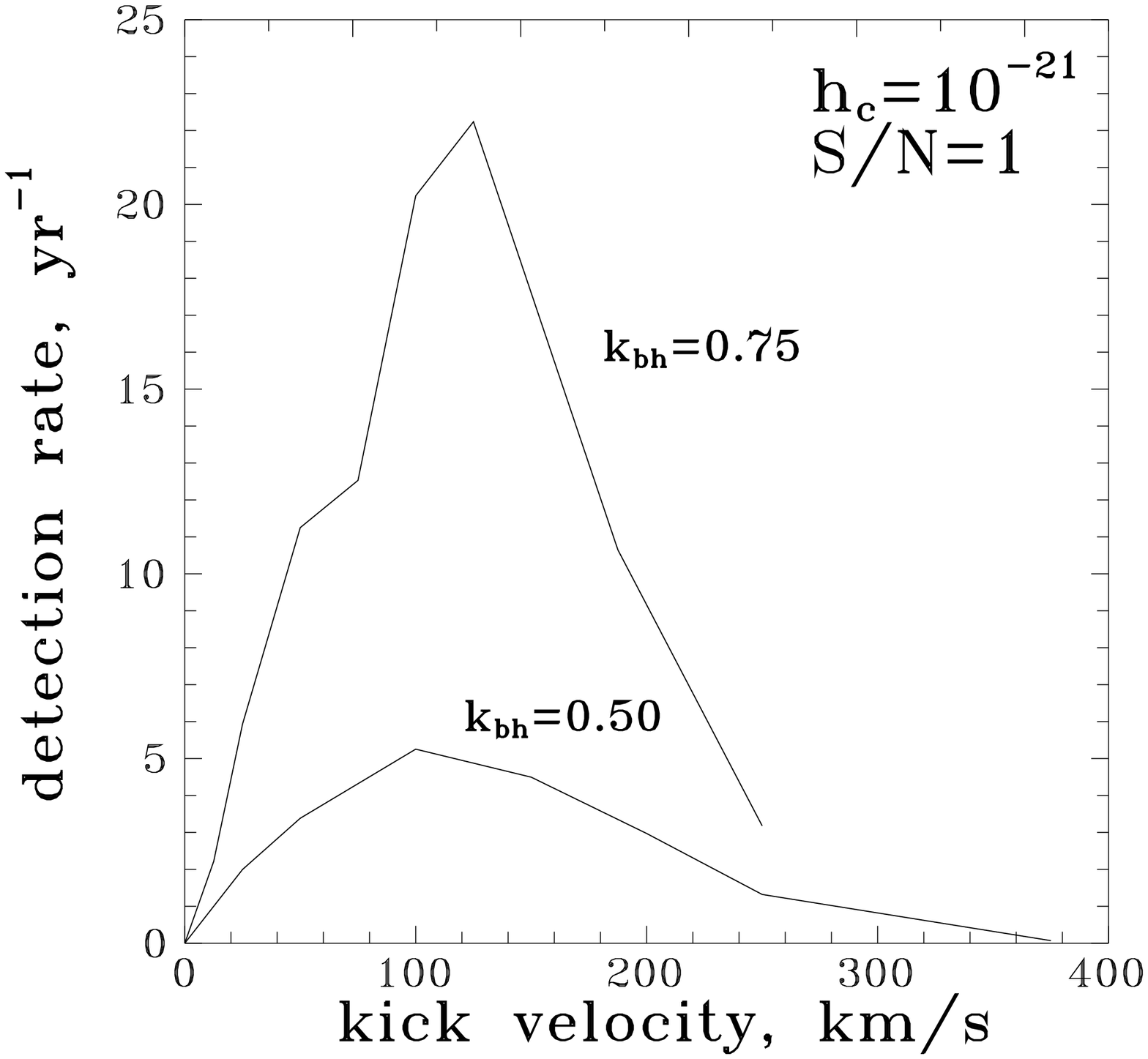,height=2.3in}
}
}
\caption{Left: BH+BH merging rates calculated for a $10^{11} M_\odot$
galaxy with a constant star formation rate, as a function of assumed kick 
velocity during BH formation with $M_{cr}=35 M_\odot$,
for $k_{bh}=0.5$ and $0.75$. \label{fig:cls-dlt}
Right: Detection rate per year of 
BH+BH mergings by the initial LIGO/VIRGO/GEO-600 laser
interferometers ($h_c=10^{-21}$ at $f=100$~Hz), as a function of kick
velocity during BH formation.\label{fig:det-dlt}}
\bigskip
\hbox to \textwidth{ 
\hbox to 0.5\textwidth{
\psfig{figure=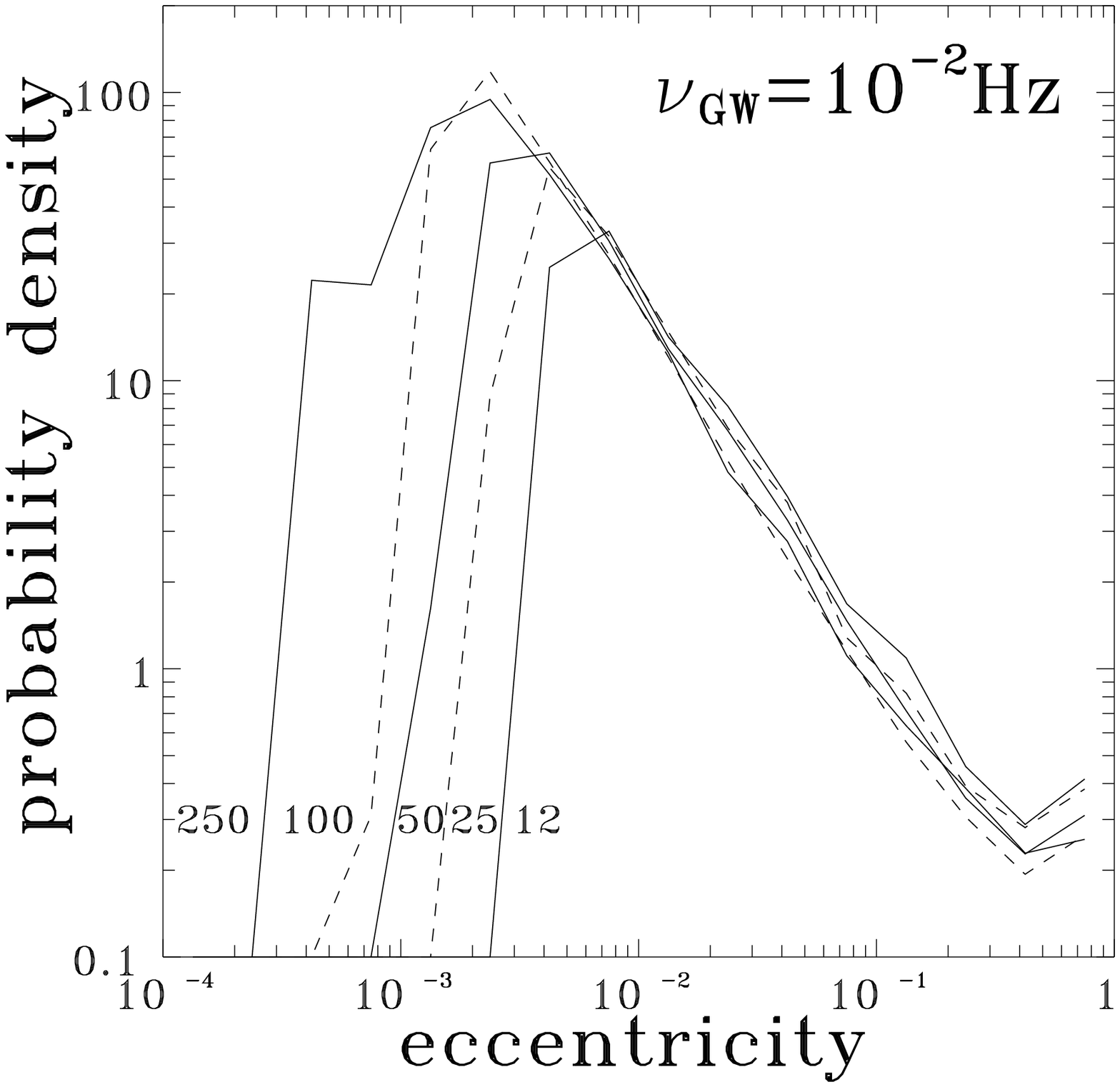,height=2.3in}
}
\hss
\hbox to 0.5\textwidth{
\psfig{figure=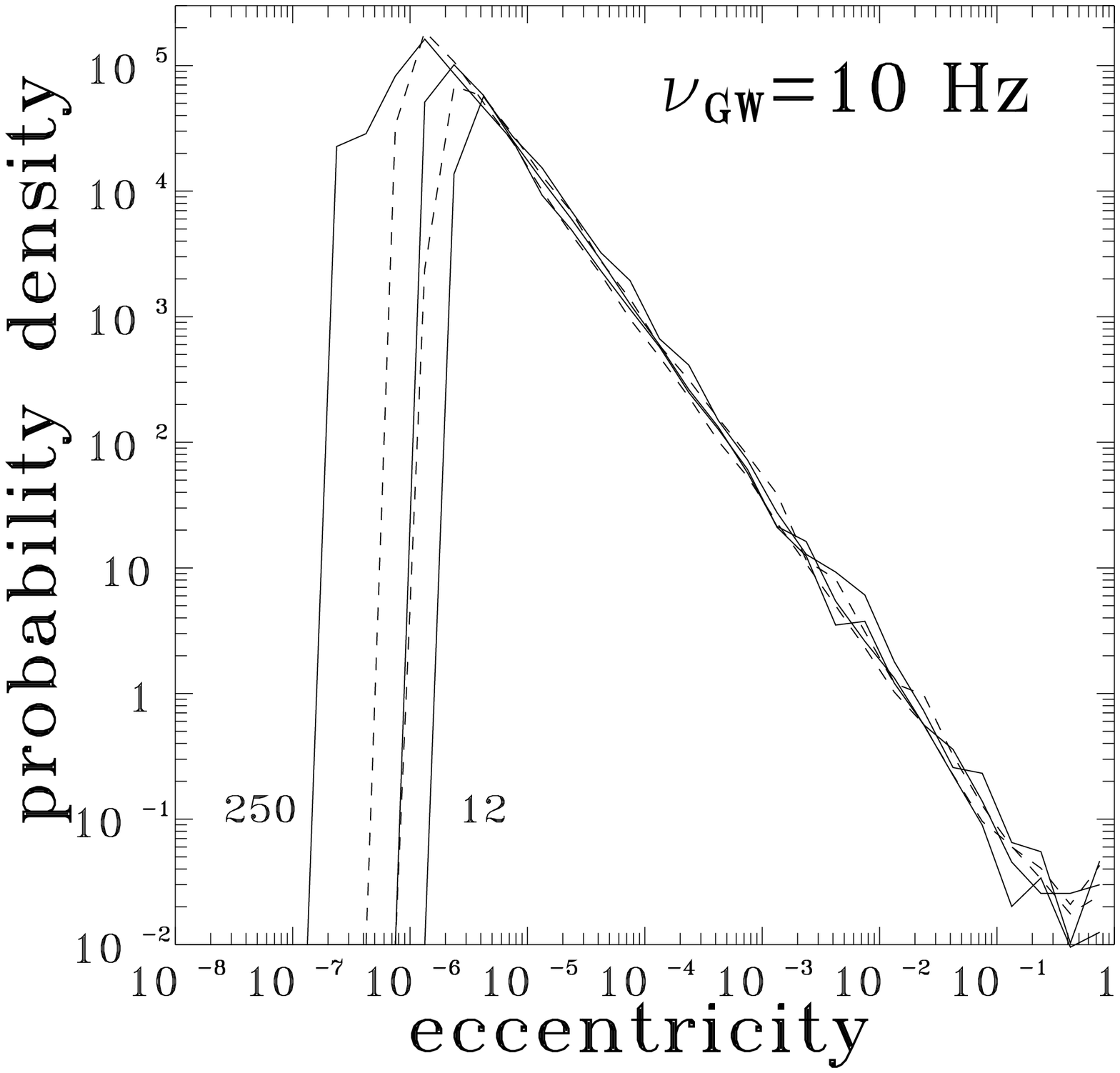,height=2.3in}
}
}
\caption{
Differential 
distribution of the orbital eccentricities of merging BH+BH systems
at frequency $10^{-2}$ Hz (left panel) and 10 Hz (right panel).
Figures mark the kick velocity amplitude.
\label{fig:ecc001}
\label{fig:ecc10}}
\bigskip
\hbox to \textwidth{\hss\psfig{figure=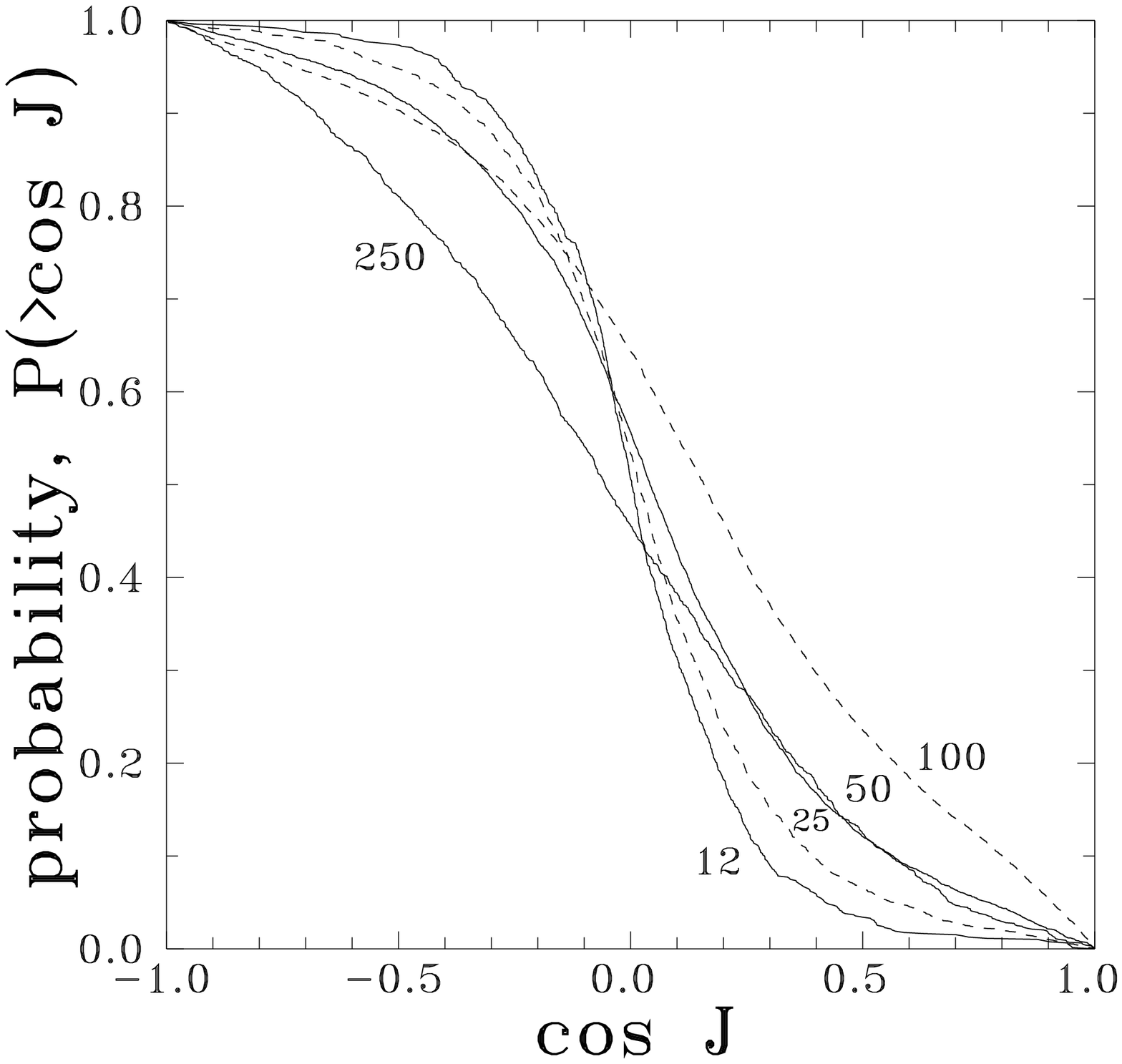,height=2.3in}\hss}
\caption{Integral distribution of the angles between the 
orbital angular momentum and BH spins 
in merging BH+BH systems. Figures mark the kick velocity amplitude.
\label{fig:icosj}}
\end{figure}

The situation, however, strongly changes for non-zero kicks during BH
formation. Then  a part of the double BH systems are 
formed vith very 
high eccentricities $e\sim 1$ after the second BH formation. As is well known 
\cite{Peters64}, the eccentricty shortens the coalescence time 
due to GW emission $t_c(e_0,\M)$ 
of a binary with the chirp mass $\M$, initial orbital period $P_0$,
and orbital eccentricity $e_0$
\begin{equation}
t_c(e_0,\M)=t_0(\M,P_0) \Phi(e_0)\,,
\label{t_c}
\end{equation}  
where 
\begin{equation}
t_0(\M,P_0)\simeq 9.8\times 10^6 [\hbox{yr}] 
\left(\frac{P_0}{1\hbox{hr}}\right)^{8/3}
\left(\frac{\M}{M_\odot}\right)^{-5/3}\,,
\label{t_0}
\end{equation}
The function 
\begin{equation}
\Phi(e_0)=\frac{48}{19}\frac{(1-e_0^2)^4}
{e_0^{48/19}\left(1+\frac{121}{304}e_0^2\right)^{3480/2299}} \int_0^{e_0}
\frac{\left(1+\frac{121}{304}e^2\right)^{1181/2299}}
{(1-e^2)^{3/2}}e^{29/19}de
\label{Phi_e}
\end{equation}
rapidly tends to zero when $e\to 1$. A high eccentricity thus enables  
binary BH systems with very large orbital semiaxes (periods) to 
merge in a Hubble time. If the kick velocity is higher than the 
orbital velocity of the exploding component, it preferentially makes 
the binary system unbind, so the dependence of binary BH merging rates on 
the kick velocity is expected to have a maximum. This effect is confirmed  
by Monte-Carlo modeling (see Fig.~\ref{fig:cls-dlt}).

\section{Population synthesis modeling of BH+BH mergings}
  
The results of population synthesis calculations of binary BH merging rate
in a model galaxy of $10^{11} M_\odot$ (assuming all stars forming in
binaries) with a constant star fromation rate are shown in Fig. 1. The
binary evolution parameters were taken in the conventional form: Salpeter
initial mass function for the primary mass $M_1$, a flat mass ratio
$q=M_2/M_1$ distribution, a flat initial semiaxes distribution $a$ (see
\cite{LPP97b} for more detail). The galactic binary BH merging rates are
shown as a function of the kick velocity during BH formation. The
calculations were done with a delta-function like kick velocity distribution
(the use of more complex distributions do not change the results
significantly and not shown in the Figure). In Fig. 1 (right panel) the
detection rate of binary BH coalescences by the initial laser
interferometers ($h_c=10^{-21}$ at $\nu=100$ Hz) in one-year integration is
shown as a function of BH kick velocity. It is seen that both the galactic
merging rate and detection rate of binary BH systems rapidly increases with
the assumed kick velocity amplitude and reach a maximum of $\R\sim 2.5\times
10^{-5}$ yr$^{-1}$ and $\D\sim 20$ detections per year, correspondingly, at
$v_k\simeq 120$ km/s. Since $\D\sim \M^{15/6}\R$, the $\R(v_k)$ and 
$\D(v_k)$ dependences have similar shapes.  

In Fig. 2 we present the distribution of eccentricities of merging BH+BH
systems at the characteristic frequencies 
$\nu_{GW}=10^{-2}$ Hz and $\nu_{GW}=10$ Hz
for LISA and LIGO/VIRGO/GEO-600 interferometers.  
The eccentricity distributions has a power-law shape 
$f(e)\sim e^{-1.3}$ in a wide range, which reflects the transformation of
the initial eccentricity in the course of 
the orbital evolution of a binary system 
of two point masses due to gravitational wave emission and the distribution 
of the intial eccentricities (i.e. immediately after the second BH 
formation) of the merging binaries. For example, for a flat initial
eccentricity distribution  we would obtain $f(e)\sim e^{-32/19}$, 
which directly follows from the relation between semimajor axis and
eccentricity evolution \cite{Peters64}. A small fraction of
merging binary BH do has appreciable eccentricities. These systems
evolve from wide, extremely eccentric binaries.    
 
Unlike a spherically symmetric
explosion of one of the components in a binary system, 
when a sudden mass loss $\Delta M$ influences only on the
binary major semiaxis and eccentricity, the asymmetric explosion with kick
velocity also changes the space orientation of the orbital angular momentum.
Unless the vector of the kick is non-central, the spin axis of the 
star is not changed. Initially, each component of a binary should have 
spins parallel to the orbital angular momentum. The spins are assumed to 
remain parallel to each other during the evolution.
After each BH formation with kick the orbital angular momentum changes its
direction, and the resulting distribution of cosines of the angle between 
the BH spins and the orbital angular momentum (denoted by $\cos J$) 
after the second BH formation 
is presented in Fig. 3. Remarkably, even for small kicks of a few ten km/s 
an appreciable fraction (30-50\%) of the merging binary BH should have 
$\cos J<0$. Note also a non-monotonic change of the distribution form with 
kick. Only a tiny fraction of binaries can have spins antiparallel to
the orbital angular momentum.  
       
\section{Conclusions}

We have carried out calculations of binary BH merging rate and 
detection rate by the initial laser GW interferometers in the framework
of the new evolutionary scenario for very massive stars
\cite{Vanbeveren_ea98}. We have found that both the merging and detection
BH+BH rates are significant from the point of view of GW detection by 
the initial LIGO/VIRGO/GEO-600 detectors only if non-zero kick velocity
is assumed during BH formation. This important point and the degree of 
compatibility of the new scenario with existing astronomical 
observations surely deserves further studies, which are in progress
\cite{LPP99}. We confirm the previous result \cite{LPP97a,LPP97b}
that binary BH still remain the most promising sources for the initial laser
GW interferometers.

Our calculations show that in this scenario, merging binary BH 
can have non-zero eccentricities at frequencies $10^{-2}$ Hz and 10 Hz 
relevant to LISA and LIGO/VIRGO/GEO-600 frequency bands. We also found that
even small kick velocities during BH formation result in an appreciable angle
between the BH spin axes and the orbital angular momentum for 
a significant (30-50\%) fraction of coalescing binary BH. These findings
should be taken into account in constructing GW templates for data
analysis from GW detectors.

\section*{Acknowledgments}
The authors acknowledge useful discussions with L.P.Grishchuk and
T.Damour.
This work is partially supported by the Royal Society Grant RCPX219.
KAP acknowledge financial support from NTP "Astronomija" (projects 
1.4.4.1 and 1.2.4.3) and RFBR grant 98-660228.

\end{document}